\documentclass[preprint,aps,prd,amsmath,showpacs,nofootinbib]{revtex4}
\usepackage{graphicx}

\newcommand{\beq}{\begin{equation}}
\newcommand{\eeq}{\end{equation}}
\newcommand{\bea}{\begin{eqnarray}}
\newcommand{\eea}{\end{eqnarray}}

\begin{document}

\title{Klein-Gordon equation for a particle in brane model}

\author{S.N. Andrianov, R.A.Daishev, S.M. Kozyrev}
\email{adrianovsn@mail.ru} \affiliation{Scientific center for
gravity wave studies ``Dulkyn'', Kazan, Russia}

\begin{abstract}
Brane model of universe is considered for free particle.
Conservation laws on the brane are obtained using the symmetry
properties of the brane. Equation of  motion is derived for a
particle using variation principle from these conservation laws.
This equation includes terms accounting the variation of brane
radius. Its solution is obtained at some approximations and
dispersion relation for a particle is derived.
\end{abstract}


\maketitle

The Klein-Gordon equation describing motion of a scalar particle
is known in quantum field theory that does not account the changes
of space metrics and changes of particles behavior connected with
it  \cite{Ryder}. These changes can be accounted by Einstein's
equation. Wheeler - de Witt equation occupies place of Einstein's
equation in quantum theory that is generalization for the case of
general relativity theory and is valid for arbitrary Ryman space
 \cite{Wheeler}, \cite{Darabi}. The approach of Wheeler - de Witt is applied to brane
theory of Universe in papers \cite{Darabi},\cite{Gusin}.

In present paper, we will derive starting from the symmetry
properties of the brane  \cite{Papantonopoulos},\cite{Langlois}
,\cite{Brax} the equation of motion for a particle in the
framework of brane model with the account of its radius variation
in universal space. This equation has a form of Klein-Gordon
equation in curved space \cite{Birrel} accounting the field of
brane fluctuations and describes particle temporal behavior with
Einstein's  or time dependent Wheeler - de Witt equation
\cite{Pavsic}.

Let's consider our space as four dimensional hyper-surface that is
the insertion in the space of higher dimension (Fig. 1). Then
interval for a moving particle in normal Gauss coordinates can be
written as ,
\begin{equation}
{\it ds}=\sqrt{g_{ij}dx^idx^j-c^2dt^2},  \label{eq1}
\end{equation}
where $g_{i j }$  is metric tensor $dx^i, dx^j$  are differentials
of coordinates (i, j = 0,1,2,3) on brane, $dt$ is differential of
universal time that is proportional to extra dimensional
coordinate. Then action can be written as

\begin{equation}
S=  mc \mathop{\int_{0}^{T}} ds =\int_0^T Ldt, \label{eq2}
\end{equation}

where $m$ is mass of particle, $c$ is speed of light,
\begin{equation}
{\it ds}=\sqrt{g_{ij}(m \stackrel{\cdot }{x}^i) (m \stackrel{\cdot
}{x}^j)-c^2dt^2}, \label{eq3}
\end{equation}
is Lagrangian, $T$ is current value of universal time in
multidimensional space (proportional to the brane radius).

Let's introduce the symmetry of configuration space as single
parametric transformation group  $f(q,\varepsilon)$:

\begin{equation}
t\rightarrow t+\varepsilon, x_i\rightarrow x_i(t+\varepsilon),
\label{e4}
\end{equation}
\begin{equation}
f(q_i,\varepsilon)= x_i(t+\varepsilon), f(q_i,0)= x_i(t)
\label{eq5}
\end{equation}
conserving Lagrangian (\ref{eq3}). According to Netter's theorem,
first integral
\begin{eqnarray}
I &=& \frac{\partial L}{\partial \stackrel{\cdot }{x}_i}h^i,
\label{eq6}
\end{eqnarray}
where
\begin{eqnarray}
h^i &=&\frac{\partial f^i}{\partial \varepsilon_{|\varepsilon=0}},
\label{eq7}
\end{eqnarray}
can be put in correspondence to each symmetry. Then

\begin{eqnarray}
I =\frac{1}{L}g_{ij}m^2 \stackrel{\cdot }{x}^j\{\frac{\partial
x_i(t+\varepsilon)}{\partial (t+\varepsilon)}\frac{\partial
(t+\varepsilon)}{\partial \varepsilon}\}_{|\varepsilon=0},
\label{eq8}
\end{eqnarray}
or
\begin{eqnarray}
g_{ij}m \stackrel{\cdot }{x}^i m \stackrel{\cdot }{x}^j =const,
\label{eq9}
\end{eqnarray}

If the particle moves uniformly and rectilinearly on the
background of Lorenz's metrics than we can choose reference system
where $\stackrel{\cdot }{x}^{(1)}=\stackrel{\cdot
}{x}^{(2)}=\stackrel{\cdot }{x}^{(3)} =0 $ and when brane is
expanding with velocity $c$. The Eq. (\ref{eq9}) yields
\begin{eqnarray}
g_{ij}m \stackrel{\cdot }{x}^i m \stackrel{\cdot }{x}^j =m^2 c^2.
\label{eq10}
\end{eqnarray}

The same equation can be derived in the framework of quantum
mechanical treatment. The wave function of particle in
quasi-classical approximation is

\begin{eqnarray}
\Psi =a e^{\frac{iS}{\hbar}}, \label{eq11}
\end{eqnarray}
where $a$ is slowly varying amplitude, $S$ is action expressed by
formulas (\ref{eq2}),(\ref{eq3}).   Let's differentiate both sides
of expression (\ref{eq11}) by $T$ neglecting the dependence of
amplitude on time
\begin{eqnarray}
\frac{d\Psi}{dt} =a
\frac{i}{\hbar}e^{\frac{iS}{\hbar}}\frac{dS}{dT}=i\frac{c}{\hbar}\sqrt{g_{ij}p^i
p^j-m^2 c^2 \Psi}, \label{eq12}
\end{eqnarray}
where  $p^{i,j}=\stackrel{\cdot }{x}^{i,j}$.

If evolution of particle does not depend on brane radius then
$\frac{d\Psi}{dt}=\frac{dS}{dt}=0$ and

\begin{eqnarray}
g_{ij}p^i p^j=m^2 c^2. \label{eq13}
\end{eqnarray}

Expression (\ref{eq13}) can be rewritten in the following form:
\begin{eqnarray}
p_i p^j=m^2 c^2. \label{eq14}
\end{eqnarray}
Let's consider functional variation \cite{Ryder} of relation
(\ref{eq14}) corresponding to brane fluctuation when coordinares
$x$ transform into coordinates  $x'$ (fig. 2). Complete variation
of momentum vector can be written as the sum of functional
variation $\delta p$ of vector  $p$  at the comparison  $p'$ of
with $p$ in the same point at the parallel transfer of vector $p$
in universal space and ordinary variation $dp$. Then, it can be
written that
\begin{equation}
\triangle p=p^{\prime }\left( x^{\prime }\right) -p\left( x\right)
=p^{\prime }\left( x^{\prime }\right) -\stackrel{\sim }{p}\left(
x^{\prime }\right) +\stackrel{\sim }{p}\left( x^{\prime }\right)
-p\left( x\right) =\delta p+dp,  \label{eq15}
\end{equation}
where
\begin{equation}
\delta p=p^{\prime }\left( x^{\prime }\right) -\ \stackrel{\sim
}{p}\left( x^{\prime }\right)  \label{eq16}
\end{equation}
and
\begin{equation}
dp=\ \stackrel{\sim }{p}\left( x^{\prime }\right) -p\left(
x\right) , \label{eq17}
\end{equation}
$\stackrel{\sim }{p}\left( x^{\prime }\right) \ $is momentum
vector at its parallel transfer in the universal space. If
trajectory of particle is geodetic one then according \cite{Ryder}

\begin{equation}
dp_i=\frac{\partial p_i}{\partial x^k}dx^k=0,  \label{eq18}
\end{equation}

\begin{equation}
\delta p_i=\stackrel{\sim }{p}_k\Gamma _{i\alpha}^k\delta
x^\alpha, \label{eq19}
\end{equation}
where $\delta x^\alpha=x^{\prime \alpha}-x^\alpha$ is variation of
brane radius. Rome indexes numerate here coordinates of usual
four-coordinate space and Greek indexes numerate coordinates of
universal five-coordinate space. It was assumed at formulation of
 (\ref{eq19}) that $\Gamma _{i\alpha}^4=0 $.

 Then, it can be written, omitting stroked index
of momentum vector,

\begin{equation}
\overrightarrow{p}\left( x^{\prime }\right)
=\overrightarrow{p}\left( x\right) +\delta \overrightarrow{p}.
\label{eq20}
\end{equation}

At the transform {\it x }$\rightarrow ${\it \ x'}, relation
(\ref{eq14}) is transforming accounting (\ref{eq20}) to the
following form:

\begin{equation}
p_ip^i+p_i\delta p^i+\delta p_ip^i+\delta p_i\delta p^i=m^2c^2.
\label{eq21}
\end{equation}

Let's pass in relation (\ref{eq21}) to operators acting in Hilbert
space of wave functions $\psi \left( x\right) $. We represent for
this sake the components of vector {\it p} as

\begin{equation}
p_i=-i\hbar \frac \partial {\partial x^i},  \label{eq22}
\end{equation}

and rewrite relation (\ref{eq19}) as

\begin{equation}
\delta p_i=-\left\{ \Gamma _{i\alpha }^k\delta x^\alpha \right\}
_{;k}, \label{eq23}
\end{equation}
assuming that $ \widetilde{p_k}$  is a covariant derivative
because of brane curvature.

Let's consider the first term in the left side of equation
(\ref{eq21}). For this purpose, we represent it in the form

\begin{equation}
p_ip^i=p_ig^{ij}p_j.  \label{eq24}
\end{equation}

Using expression (\ref{eq22}), we get

\begin{equation}
p_ip^i=-\hbar ^2\left( \frac{\partial g^{ij}}{\partial x^i}\frac
\partial {\partial x^j}+g^{ij}\frac{\partial ^2}{\partial
x^i\partial x^j}\right) . \label{eq25}
\end{equation}

Let's use well known relation

\begin{equation}
\frac{\partial g^{ij}}{\partial x^k}=-\Gamma _{mk}^ig^{mj}-\Gamma
_{mk}^jg^{im}.  \label{eq26}
\end{equation}

Then

\begin{equation}
p_ip^i=-\hbar ^2\left( g^{ij}\frac{\partial ^2}{\partial x^i\partial x^j}%
-g^{mj}\Gamma _{mi}^i\frac \partial {\partial x^j}-g^{im}\Gamma
_{mi}^j\frac
\partial {\partial x^i}\right) .  \label{eq27}
\end{equation}
Changing indexes of summation, we get

\begin{equation}
p_ip^i=-\hbar ^2g^{ij}\left( \frac{\partial ^2}{\partial x^i\partial x^j}%
-\Gamma _{ik}^k\frac \partial {\partial x^j}-\Gamma _{ij}^k\frac
\partial {\partial x^k}\right) .  \label{eq28}
\end{equation}
Let's consider second term in the left side of equation
(\ref{eq15}), rewriting it in the form

\begin{equation}
p_i\delta p^i=p_ig^{ij}\delta p_j.  \label{eq29}
\end{equation}

Using formula (\ref{eq26}), we get

\begin{equation}
p_i\delta p^i=g^{ij}p_i\delta p_j+i\hbar \left( g^{ij}\Gamma
_{im}^m+g^{im}\Gamma _{im}^j\right) \delta p_j + g^{ij}\delta p_j
p_i. \label{eq30}
\end{equation}

Let's write in its direct form the covariant derivative in the expression (%
\ref{eq23}):

\begin{equation}
\delta p_j=-i\hbar \left( \Gamma _{jk}^k +\frac{\partial \Gamma _{j\alpha}^k}{%
\partial x^k}\delta x^\alpha-\Gamma _{lk}^k\Gamma _{l\alpha}^k\delta x^\alpha+\Gamma
_{lk}^k\Gamma _{j\alpha}^l\delta x^\alpha\right) ,  \label{eq31}
\end{equation}

where stroked index of the derivative on is omitted.We get from
formula \ref{eq31})
\begin{equation}
\delta p_j=-i\hbar \left( \Gamma _{jk}^k+R_{j \alpha}\delta x^\alpha+\frac{\partial \Gamma _{jk}^k}{%
\partial x^\alpha}\delta x^\alpha\right) ,  \label{eq32}
\end{equation}
Substituting expression (\ref{eq31}) into formula (\ref{eq30}), we
get

\begin{eqnarray}
&&
\begin{tabular}{l}
$p_i\delta p^i=-\hbar ^2g^{ij}\left(\frac{\partial
R_{j\alpha}}{\partial x^i} \delta x^\alpha -R_{ij} -\Gamma _{mi}^m
\Gamma_{jk}^k - \Gamma_{mi}^m R_{j\alpha}\delta
x^\alpha-\Gamma_{ij}^l
\Gamma_{lk}^k - \Gamma_{ij}^l R_{l\alpha}\delta x^\alpha\right) +$ \\
\\
$+g^{ij}\delta p_j p_i$
\end{tabular}
\label{eq33}
\end{eqnarray}

\begin{equation}
\delta p_i p^i=-\hbar ^2g^{ij} \left( R_{i\alpha}\delta
x^\alpha-\Gamma_{ik}^k\right)\frac{\partial }{\partial x^j}
\label{eq34}
\end{equation}`

where $ \delta p_l=0 $ was assumed after taking the derivatives.
From now up to the end of paper, we will denote by $\alpha$  only
the extra dimensional coordinate. Here,  $ \frac{\partial \Gamma
_{ik}^k}{\partial x^\alpha}=0 $   was assumed. Apparently,

\begin{equation}
\delta p_i \delta p^i=-\hbar ^2g^{ij} \left( \Gamma _{im}^m
\Gamma_{jk}^k + \Gamma_{ik}^k R_{j\alpha}\delta x^\alpha+
\Gamma_{jk}^k R_{i\alpha}\delta x^\alpha + R_{i\alpha} R_{j
{\alpha ^ \prime}} \delta x^\alpha \delta x^{\alpha ^\prime}
\right) \label{eq35}
\end{equation}`

 Using equations (\ref{eq21}, \ref{eq28}, \ref{eq33}, \ref{eq34}, \ref{eq35}), we get

\begin{eqnarray}
&&
\begin{tabular}{l}
$\hbar ^2g^{ij}(\frac{\partial ^2}{\partial x^i\partial x^j}
 -\Gamma _{ij}^k \frac{\partial }{\partial x^k}+\Gamma _{ik}^k \frac{\partial }{\partial x^j}
-\Gamma_{ij}^l\Gamma_{lk}^k + \frac{\partial R_{j\alpha}}{\partial
x^i}\delta x^\alpha -
\Gamma_{ij}^l R_{l\alpha}\delta x^\alpha + 2 R_{i \alpha} \delta x^\alpha \frac{\partial }{\partial x^l}+$ \\
\\
$+\Gamma_{ik}^k R_{j\alpha}\delta x^\alpha + R_{i\alpha} R_{j
{\alpha ^ \prime}} \delta x^\alpha \delta x^{\alpha
^\prime}-R_{ij}) \psi^n+m^2 c^2\psi^n=0$
\end{tabular}
\label{eq36}
\end{eqnarray}

The first order covariant derivative on $\psi^n $  is

\begin{equation}
\{ \psi ^n\}_{;i}=\frac{\partial \psi ^n}{\partial
x^i}+\Gamma_{ik}^n \psi^k \label{eq37}
\end{equation}
 Let's denote the "geometrical" part of partial wave function derivative as
\begin{equation}
(\psi ^n)^\prime=\Gamma_{ik}^n \psi^k \label{eq39}
\end{equation}
Then we can write using orthogonal character of wave functions
\begin{equation}
(\psi ^n)^\prime=\psi ^n \mathop{\sum_{k}}\psi_k( \psi
^k)^\prime=\psi ^n \mathop{\sum_{k}}\psi_k \mathop{\sum_{l}}
\Gamma_{il}^k \psi^l=\mathop{\sum_{k}} \Gamma_{ik}^k \psi^n
\label{eq40}
\end{equation}

and
\begin{equation}
\Gamma_{ik}^n \psi^k= \Gamma_{ik}^k \psi^n \label{eq41}
\end{equation}
If wave function is a scalar  $(\psi ^n)^\prime=\Gamma_{ik}^n
\psi^k $. Hence, $\Gamma_{ik}^k=0$
Then we can rewrite (\ref{eq36}) as
\begin{equation}
\{ g^{ij} \left(D_i +\delta x^\alpha  R_{i \alpha}
\right)\left(D_j +\delta x^\alpha  R_{j \alpha} \right)+
\left(\frac{m c}{\hbar }\right)^2\}\psi=R\psi, \label{eq42}
\end{equation}
where $ D_i, D_j$  are covariant derivatives and $R$ is scalar
curvature.

\begin{figure}
\begin{center}
\includegraphics[width=7cm]{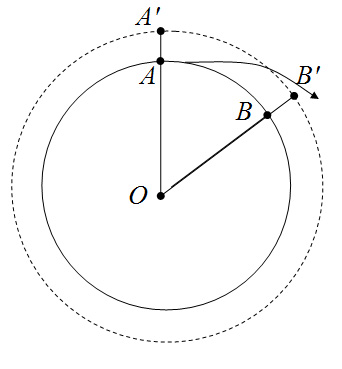}
\end{center}
\caption{Brane model of Universe. Motion of particle from point A
to point B(B') in expanding Universe $(AB=\nu d \tau,  AB'=c d
\tau, BB'= c dt )$. }
\end{figure}

\begin{figure}
\begin{center}
 \includegraphics[width=7cm]{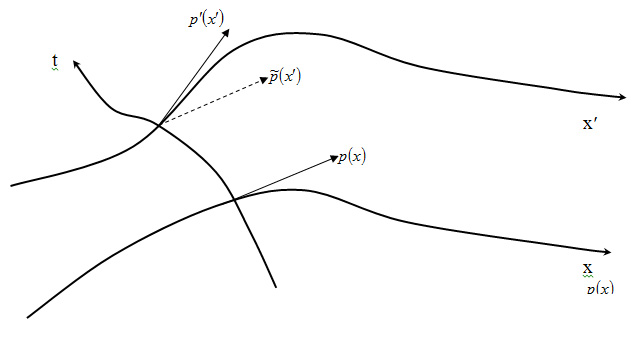}
\end{center}
\caption{Coordinate transform at the excitation of membrane. }
\end{figure}

\begin{figure}
\begin{center}
\includegraphics[width=7cm]{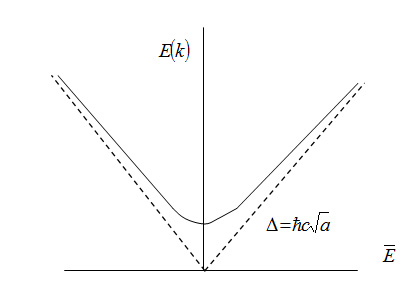}
\end{center}
\caption{Energy of moving particle $E(k)=ck$. }
\end{figure}

Let's consider small region of space-time where we can suppose
gravitation field to be constant and homogeneous. We rewrite
equation (\ref{eq43}) in locally-geodesic coordinate system for
particle in the following form:
\begin{equation}
\left\{ \frac{\partial^2 }{\partial x^2} + \gamma \frac{\partial
}{\partial x } + a\right\} \psi = \frac{1 }{c^2} \frac{\partial^2
\psi}{\partial t^2} \label{eq43}
\end{equation}
limiting ourselves by one spatial dimension, assuming the absence
of affine connection in time and introducing the notations

\begin{equation}
 \gamma =2 R_{1 \alpha}\delta x^\alpha
  \label{eq44}
\end{equation}
and
\begin{equation}
a=R - \left(\frac{m c}{\hbar }\right)^2 - \left(\frac{\partial
}{\partial x}   R_{1 \alpha}\right)\delta x^\alpha -  R_{1 \alpha}
R_{1 \alpha} \delta x^\alpha \delta x^\alpha,
 \label{eq45}
\end{equation}
Let's look for solution in the form
\begin{equation}
\psi=e^{i(kx-\omega t)},
 \label{eq46}
\end{equation}

Then we get from (\ref{eq43}) the characteristic equation

\begin{equation}
k^2 -i \gamma k  - \left(a+\frac{\omega^2 }{c^2} \right)=0,
 \label{eq47}
\end{equation}
Its physically reasonable solution is
\begin{equation}
k= \sqrt{\frac{\omega^2 }{c^2}+a-\frac{\gamma^2 }{4}} +i
\frac{\gamma }{2},
 \label{eq48}
\end{equation}
Apparently it is valid when $a$ does not depend on coordinate and
time. When $a=\gamma=0 $ , we have the usual dispersion relation
$\omega_0=ck $.

When $ \left(\frac{\omega }{c}\right) \gg a- \left(\frac{\gamma
}{2}\right)^2 $  we can approximately get
\begin{equation}
k= \frac{\omega }{c }\left(1+ \frac{c^2}{\omega^2 }\left(
a-\frac{\gamma^2 }{4}\right)\right)-i \frac{\gamma}{2},
 \label{eq49}
\end{equation}

and for phase velocity of mass-less particles
\begin{equation} \frac{dk}{d\omega}= c \left(1+
\frac{c^2}{\omega^2 }\left( a-\frac{\gamma^2 }{4}\right)\right),
 \label{eq50}
\end{equation}

Formula (\ref{eq50}) shows that effective refractive index related
with space curvature is equal

\begin{equation}
n_{eff}= 1- \frac{c^2}{\omega^2 }\left( a-\frac{\gamma^2
}{4}\right),
 \label{eq51}
\end{equation}
Thus, the phase velocity of mass-less particles in universal space
can exceed the phase velocity of light in plane Deckard space
because of drift of particles at the expansion of Universe. Other
consequences of space curvature are the following two facts that
realize when expression (\ref{eq48}) is valid:

   - it is impossible to create particle with kinetic energy
less than $\hbar \sqrt{\omega^2+ac^2} $ (fig. 3);

- space curvature leads to the frequency shift according the
formula $\omega=\sqrt{\omega_0^2-ac^2}$ that gives the possibility
for verification of developed model when curvature varies at the
influence brane fluctuations in universal space.

\section*{Acknowledgments}\nonumber
We are grateful to  V.V. Bochkarev and V.Kurbanova for the helpful
discussions.

\end{document}